\documentstyle [12pt] {article}
\def\double{\baselineskip 24pt \lineskip 10pt}

\begin{document}
\centerline{\large\bf Numerical Experiments on String Cosmology}
\vskip24pt
\baselineskip=10pt
\centerline{Mairi Sakellariadou
\footnote{Current address: Universit\'e de Gen\`eve, D\'epartement
de Physique Th\'eorique,
24 quai Ernest Anserment, CH--1211 Gen\`eve, Switzerland;
e--mail address: mairi@karystos.unige.ch}     }
\vskip10pt
\centerline{\it Institut f\"ur Theoretische Physik}
\centerline{\it der Universit\"at Z\"urich, Winterthurerstr. 190,}
\centerline{\it CH-8057 Z\"urich, Switzerland}
\vskip 44pt
\begin{abstract}
\vskip 10pt
We investigate some classical aspects of fundamental strings via numerical
experiments.  In particular, we study the thermodynamics of a string
network within a toroidal universe, as a function of string energy density
and space dimensionality.  We  find that when the energy density of the system
is low, the dominant part of the string is in the form of closed loops of the
shortest allowed size, which correspond to the momentum string modes.  At a
certain critical energy density corresponding to the Hagedorn temperature,
the system
undergoes a phase transition characterized by the formation of very long loops,
winding a number of times around the torus.  These loops correspond to the
winding string modes.  As the energy density is increased, all the extra energy
goes into these long strings.  We then study the lifetime of winding modes as
a function of the space densionality.  We find that in the low--energy
density regime, long winding strings decay only if the space dimensionality
of the toroidal universe is equal to 3.  This finding supports the proposed
cosmological scenario by Brandenberger and Vafa, which attempts to explain the
space dimensionality and to avoid the initial singularity by means of string
theory.
\vspace{1cm}
\end{abstract}
\vspace{1cm}

\double

\section{Introduction}

One of the main questions facing string theory in many potentially
exciting applications is the understanding of the behaviour
of strings in extreme conditions of high temperatures and high
densities.  One would expect these domains to be
achieved in the context of the eraly universe.
If string theory is a theory unifying matter and gravity,
then one should expect  the evolution of the universe,
in particular around the Planck scale, to deviate from
its description according to standard cosmology.
Our homogeneous, isotropic and almost flat universe, should
be the remnant of the evolution of the universe as described by
string theory. Thus we need to do
string cosmology, in other words, to study string behaviour
at high energy density and temperature.
In addition, since stringy effects become important at an
energy scale much beyond the one we can probe in high energy
scattering experiments, obtaining currently measurable
cosmological consequences of string theory will allow us to
get a confrontation between string theory and experiments.

However the main difficulty in using string theory in such a context
lies in the fact that one would have to know about non--perturbative
aspects of strings, which we do not have much understanding of
at the present.  It would thus be desirable to have a model for strings
which one can trust in such domains.  In this paper we explore one
such model and its physical implications.

Usually when we discuss string theory we immediately turn to the
quantum theory.  We consider first quantized strings consisting
of the graviton and all the infinitely many massive stringy excitations
and talk about their interactions etc.  This was indeed the main motivation
for studying string theory in the first place, {\it i.e.,} the fact that it
leads to a consistent {\it quantum} theory of gravity.  However,
one expects that many properties of strings should be dominated by
{\it classical} aspects of strings.  In this context one could deal
with all the modes of strings at once which are in a coherent state
and behave in a classical way.  Our aim is to study some classical aspects
that strings have.  There are two postulates we are going to make
which we find reasonable:

1--Classical strings are simply described by the Nambu \cite{n} -- Goto
 \cite{g} action
and thus behave very much like cosmic strings.  Moreover
it will be assumed that whenever two strings meet at a point they
intercommute with a certain probability.

2--There is a maximum energy density for strings which one can identify with
the Planck density.

The first postulate is more or less the definition of classical
strings modulo the condition for intercommutation when they meet,
which is there to mimick string interactions.  The second postulate
is intuitively what one would expect but is harder to give a rigorous
justification for.  In a sense it is in accord with the intuition
that there can be at most one string per Planck volume and that we cannot
put infinitely many strings squeezed into a small region of space.
At any rate this will be our assumption in this paper.  Note
that a particular consequence of this assumption is that the only
degree of freedom for strings
with the maximum density is how they are connected.  The connectivity
of strings becomes the relevant question near this maximum density.

Luckily both of these postulates are realized in the context of cosmic string
simulations on a lattice \cite{shell} ,\cite{alex},
which one can thus use in the context of fundamental strings.

Here, the aim is to study via numerical experiments the
thermal properties of strings and equilibration
process for strings, and in particular its dimension
dependence.  One motivation for this study is to test the
model proposed in Ref.~\cite{br-v} for string cosmology.
That model, which takes the space to be a $D$--dimensional
torus, avoids the singularity of Friedmann--Robertson--Walker cosmology by
a stringy method which is the well known $L\mapsto 1/L$
duality of strings \cite{dual}, stating that large and small tori
are equivalent.  It was further suggested that the dynamics
dictating the evolution of the universe should be
modified in the context of string theory
from Einstein's equations in order for it to be
consistent with this small--large duality.   Moreover,
it was suggested there that if long strings do not
decay, then the expansion of the universe may stop
as it would otherwise create perpetual energy.
This latter point, which was postulated there,
together with the modification needed to bring
Einstein's equation in line with stringy duality,
was shown \cite{ts-v}
to be the consequence of the existence of the dilaton
in string theory.  It was further suggested in Ref.~\cite{br-v} that
the macroscopic dimension of space may be 3, because long
strings in higher dimensions have a harder time chopping themselves
up to short strings, which combined with the equations of string
evolution will be consistent with expansion only for space
dimensions less than or equal to three\footnote{The dimension 3 is the most
likely direction to expand because it has the larger phase space.}.
The main open question here is whether or not long strings
survive unusually long in higher dimensions.  This will be put
here to the test in the context of numerical simulations of strings.
As a check on the numerical simulation and its equivalence
to quantum strings, we compare the distribution of strings in
thermal equilibrium studied for quantum strings in Refs.~\cite{bw} --
\cite{jain}, with the numerical results as a function of space dimension.

This article is organized as follows: in section 2 we descibe our
`experimental' setup (our numerical algorithm); in section 3 we
study the thermodynamics of a string network as a function of string
energy density and space dimensionality---we first explain what
the prediction of quantum strings is and then compare it to what
we find numerically; in section 4 we examine
whether the lifetime of long strings depends on the space dimensionality;
and in section 5 we state our conclusions.

\section{ Description of the Model}
\vskip 0.5 true cm
Before describing our discrete model we will briefly discuss the
dynamics of continuous cosmic strings. Strings are represented
 \cite{old}
by a vector function $\vec{f}(\sigma , t)$, which satisfies
\begin{equation}
\vec{f}^{\ \prime}\cdot \dot{\vec{f}}\ \ =\ \ 0\ \ ;\ \
\vec {f}^{\ \prime 2}\ +\ \dot{\vec{f}^2}\ \ =\ \ 1\ , \label{gauge}
\end{equation}
where $\sigma $ is a parameter along the string; primes
and overdots denote derivatives with respect to $\sigma$ and $t$,
respectively. Equations (\ref{gauge}) are gauge conditions on
the parametrization of the string worldsheet, so that the Nambu -- Goto
equation of motion in flat spacetime reduces to the wave equation
\begin{equation}
\ddot{\vec{f}}\ \ -\ \ \vec{f}^{\ \prime\prime}\ \  =\ \ 0\ . \label{wave}
\end{equation}
(The Nambu--Goto action leading to the above equation is a good
approximation for the motion of cosmic strings as long as their
radius of curvature is larger than their width, but it is exact
as far as classical fundamental strings are concerned.)
The meaning of those  gauge conditions is that a string
segment $d\sigma $ moves perpendicular to itself, carrying an
energy  $\mu d\sigma $ ($\mu$ denotes the mass per unit length of a
 string).
Equation (\ref{wave}) admits the general solution
\begin{equation}
\vec{f}(\sigma ,t)\ \ =\ \ {1\over 2}\  [ \vec {a}
(\sigma - t) + \vec{b}(\sigma +t)]\ ,\label{solwave}
\end{equation}
where $\vec a$ and $\vec b$ are arbitrary vectors on a unit sphere.
As strings move they may exchange partners at intersection points, with
an intercommuting probability depending on the string coupling\footnote{
In string theory the probability may change depending on the value
of the dilaton.}.

We will now describe the model we will use.  It is
based on an algorithm proposed
by Smith and Vilenkin \cite{alex}
to which we refer the interested reader for more details.
Here we will briefly mention some of its main properties.
In our `experimental' setup, each string is represented by
a discrete number of
points on it, which are equally spaced in the parameter $\sigma $
by a separation  $\delta $. We consider
the segments of strings between neighbouring points to have all the same
energy. In other words, points on strings
are equally spaced in energy, meaning that the total energy of a
string is proportional to the number of points by which it is
represented. We choose as unit of energy  the energy between
neighbouring string points, while as unit of length and time
the lattice spacing $\delta $. Strings are contained in a
$D$--dimensional box of size $L$ with periodic boundary conditions, {\it i.e.,}
a  torus. (Each string leaving the box reenters through the opposite face.)
To evolve the sequence of points by which we are representing strings,
we define \cite{alex}
\begin{equation}
\vec{v}(\sigma,t)\ \ \equiv \ \ {1\over\delta}
[\vec{f}(\sigma,t+\delta)-{1\over 2}[\vec{f}(\sigma+\delta,t)+
\vec{f}(\sigma-\delta,t)]]\ ,\label{vel}
\end{equation}
as the discrete version of the string velocity $\dot{\vec{f}}
(\sigma,t)$, becoming exact as $\delta\rightarrow 0$.
Provided we know $\vec{f}$ and $\vec{v}$ at one instant,
the equations
\begin{equation}
\vec{f}(\sigma,t+\delta)\ \ =\
\ {1\over 2}[\vec{f}(\sigma+\delta,t)+\vec
{f}(\sigma -\delta,t)]+\vec{v}(\sigma,t)\delta \ ,\label{evi}
\end{equation}
and
\begin{equation}
\vec{v}(\sigma,t+\delta)\ =\ {1\over 2}[\vec{v}(\sigma+\delta,t)+
\vec{v}(\sigma-\delta,t)]+{1\over 4\delta}[\vec{f}(\sigma+2\delta,t)-
2\vec{f}(\sigma,t)+\vec{f}(\sigma-2\delta,t)]\ ,\label{evii}
\end{equation}
enable us to calculate both $\vec{f}$ and $\vec{v}$ at all times;
Eqs.~(\ref{evi}), (\ref{evii}) create an exact solution of Eq.~(\ref{wave}).
($\vec{f}$ and $\vec{v}$
are determined on two interlocking but non--overlapping
diamond lattices in the $(\sigma ,t)$ plane.) Two consecutive
points on a string segment determine a link of energy $2\mu\delta$
moving at velocity $\vec{v}$. In this representation we assign
position (but not velocity) to the points on the strings, while
we assign  velocity (but not position) to the links joining the
string points.
The discreteness of our model imposes a lower energy cutoff;
we choose the cutoff parameter $E_c$ to be equal to $2$, so that the shortest
loops have just two links. Defining
\begin{equation}
\vec{u}(\sigma ,t)\ \ \equiv\ \ {1\over 2\delta} [\vec{f}
(\sigma +\delta,t)-\vec{f}(\sigma -\delta,t)]\ ,\label{uu}
\end{equation}
as the discrete version of $\vec{f}^{\ \prime}$,
the gauge conditions, Eq.~(\ref{gauge}), take the discrete form
\begin{equation}
\vec{u}\ \cdot\ \vec{v}\ \ =\ \ 0\ ;
\ \vec{u}^2\ +\ \vec{v}^2\ \ =\ \ 1\  \label{disg}
\end{equation}
and are preserved by the discrete evolution equations (Eqs.~(\ref{evi}),
(\ref{evii})).
We evolve all strings ahead by one timestep $\delta $
and subsequently allow them to intercommute, namely two strings passing through
the same lattice site  simply reconnect with probability $p$.
We choose $p=1$ in this paper (but it would  also be interesting
for future work to allow this to vary with the value of the dilaton).
 We then  proceed with the next timestep.
An attractive feature of our model is that the
algorithm for the string dynamics exactly integrates the equations
of motion, while the intercommuting algorithm exactly preserves
energy and momentum.

In this experimental setup a string state of energy $E$ will
look like a random walk consisting of $E$ points on the lattice,
which will have a physical length scaling like $\sqrt{E}$.
In our experimental setup it is convenient to divide string
states in two types: the {\it long} ones and the {\it short} ones.
To study string thermodynamics we classify a loop as a short one
if its energy is less than $E<L^2$ and long one if its energy is
equal or greater than $E>L^2$. Note that since a typical configuration
of a string is a random walk a long loop typically winds at least once
around the toroidal universe.  We have checked that the results we find
are insensitive to the precise choice of the definition of short and long
strings.  To study the lifetime of the winding string modes, we employ a
more restrictive criterion.  At each timestep of the evolution
of the string network, we calculate the extent of all loops in all
directions.  A loop will be defined as a long one if at least one of its
extents exceeds $kL$, where $k$ is a numerical coefficent and $L$ stands
for the size of the $D$--dimensional toroidal box.

\section{String Thermodynamics}

It is one of the basic facts about string theory that the degeneracy
of string states increases exponentially
with energy
\begin{equation}
d(E)\ \sim \ \ \exp(\beta_H E)\ . \label{ener}
\end{equation}
A simple consequence of this is that there is a maximum
temperature which is $T_{max}=\beta_H^{-1}$, known
as the Hagedorn temperature \cite{hag}, \cite{carl}, \cite{fr}.
In the microcanonical ensemble the description of this situation is
as follows: We put strings in a box with a certain total energy.
Let us consider the energy density
\begin{equation}
\rho = {E\over L^D}\nonumber\ .
\end{equation}
Now for energies small enough, or more
precisely for $\rho \ll 1$, the configuration of strings in
equilibrium will be dominated by short strings ({\it i.e.,} the massless
modes in the quantum description).  However as we increase the energy
density more and more oscillatory modes of strings get excited.
In particular, as shown in the work of Refs.~\cite{bw} --  \cite{jain},
if we reach a critical density $\rho_H$ then long oscillatory string
states begin to appear in the equilibrium state.  The density
at which this happens corresponds to the Hagedorn temperature.
  For any density
$\rho >\rho_H$ the long oscillatory strings continue to
be present in the equilibrium state and take up more
and more of the total energy available.
The description of the number of strings of various string energies
for $\rho >\rho_H$
was worked out in Ref.~\cite{jain} for fundamental strings in a box.  The
conclusion is as follows:  There is a clear separation of string
states which are long and oscillatory and the ones which are short.
For short strings the number distribution in a given energy range
goes as
\begin{equation}
{dn\over dE}\propto E^{-(1+D/2)}\ . \label{shortd}
\end{equation}
Note that this means that there is a {\it finite} amount of total energy
that the short strings can have.  Let us denote this energy by $E_H=
\rho_H L^D$.
 So all the excess energy $E-E_H$
goes to the production of long strings.  The number distribution for
long strings goes as
\begin{equation}
{dn\over dE}={1\over E}\ ,\label{longd}
\end{equation}
which means than the total number of long strings is roughly
${\rm log}(E-E_H)$.  So typically the number of long strings
grows very slowly with energy (in particular in our simulations
we find that we only have typically a few of them at
equilibrium).  So in a situation with $\rho >\rho_H$ we have
a few long strings which take up most of the energy of the system.
The small loops behave like a `background gas'.

There is an order parameter for the Hagedorn phase transition:
Let $\tau_l(E)$ denote the lifetime of a long string
of energy $E$; by lifetime we mean the time it takes
for the long string to decay into short string states.
Then for
$\rho < \rho_H$ all the long strings decay and disappear
from the thermal equilibrium state so that $\tau_l <\infty$.  However
for $\rho\geq \rho_H$ since long strings persist at equilibrium
we have $\tau_l =\infty$.  In particular, this means that there
is an exponent $\alpha (E)$
that captures this transition,
namely if we approach $\rho \rightarrow \rho_H$
from below, then
\begin{equation}
\tau_l(E)\propto {\rho_H -\rho \over \rho_H}^{-\alpha(E)}\ .
\end{equation}

All of this is well defined for quantum string theory.  However
it would be nice if these features were also true in the classical
domain of strings.  Indeed, we will now see that the lattice
model we described in the previous section has exactly the same features
as expected for fundamental strings.  The case of $D=3$ has
already been studied in Ref.~\cite{mairi1}, which indeed
shares the same thermodynamical properties as we described above.
So the main aim of our simulations is twofold:

1--To check this
correspondence for higher $D$ and in particular compute $\rho_H$ as
a function of $D$.  This is not a trivial statement and {\it a priori}
it may have not been clear that string reconnections in this lattice
model we are considering are frequent enough to even reach equilibrium.
In particular, there is the question that,  since the strings may have a hard
time finding one another, there may be no dynamical way of reaching
equilibrium in string theory in higher dimensions.

2--To Study the lifetime of long strings
and its dependence on dimension.  As discussed in the introduction,
this issue was motivated by questions raised by string cosmology.

As a first step we would like to check whether the existence
of a phase transition characterized by the appearenace of
long loops is a generic property independent of the space
dimensionality $D$. Secondly, we want to study the dependence
of $\rho_H$ on the space dimensionality $D$. Since we expect that
any dependence on $D$ will appear once
we go from $D=3$ to $D=4$, and also due to computer limitations,
we investigate how the equilibrium properties of a string
system depend on the space dimensionality $D$, for $D=3, 4$, or $5$.

To study the equilibrium properties of the string network we
perform computer simulations at various string densities. The
initial states of those runs were chosen in the form of a `loop
gas', consisting of the shortest  loops with randomly
assigned positions and velocities. Each of these loops has
just two `degenerate' links. A degenerate link has \cite{alex}
$\vec {u} = 0$ and
$|\vec {v} | = 1$  (one component of $\vec{v}$ is $\pm 1$ and the
other two zero); its end points are degenerate, $\Delta \vec {f} = 0$.
Such links are fully contracted and  move at speed of light; they are a
discrete version of the cusps that exist in continuous strings.
The state of equilibrium should
not depend on  the choice of initial state. The loop gas was
chosen both because its energy density is easily adjustable,
and also because the Monte Carlo algorithm simulating the formation
of strings at a phase transition \cite{vavi} is only applicable for
$D=3$. We ran our simulations
in $D$--dimensional boxes of size $L$ from $10$ to $50$ for $D=3$;
and $L$ from $10$ to $20$ for $D=4, 5$.

Since, as we will discuss, we find that in accordance with the
prediction of fundamental string equilibrium, there is a fixed maximum
energy going to the creation of short loops for $\rho >\rho_H$, in order
to compute the critical Hagedorn
density $\rho_H$ we can consider {\it any} thermal
equilibrium with $\rho > \rho_H$ and find $\rho_H$ from
\begin{equation}
\rho_H\ \ = \ \ \rho \ \ x_s \ ,\label{crit}
\end{equation}
where $x_s$ denotes the ratio of lattice points
occupied by short loops to the number of lattice
points occupied by loops of all sizes
(which is $N$) at equilibrium. (Note that $\rho \equiv N/L^D$.)
The fact that $\rho_H$ thus obtained does not depend on which
$\rho>\rho_H$ we start with, is a confirmation of the fact
that beyond Hagedorn density all the energy goes
to the creation of long strings.

We have indeed found a sharp change of behaviour at $\rho \sim \rho_H$.
For $\rho <\rho_H$, there are no long loops, their
energy density, $\rho_l$, is zero. For $\rho >
\rho_H$, the energy density in short loops, $\rho_s$, is constant
and equal to $\rho_H$, while
\begin{equation}
\rho_l\ \ = \ \ \rho\ - \ \rho_H\ \ \
(for \ \rho >  \rho_H)\ .\label{anoth}
\end{equation}
Figure 1 shows the energy density in short loops, $\rho_s=\rho -\rho_l$,
as a function of the string density $\rho$.
The value of $\rho_H$ depends on the size of the smallest allowed loops.
If the smallest allowed loops consist of only two points, then the values we
obtain for the critical string density, performing runs
at rather high string densities, are:
\begin{eqnarray}
\rho_H\ & =
\cases{\ 0.172 \ \pm \ 0.002 \ & for \  $D=3$\ ;\cr
       \ 0.062 \ \pm \ 0.001 \ & for \  $D=4$\ ;\cr
       \ 0.031 \ \pm \ 0.001 \ & for \  $D=5$\ .\cr} \label{critic}
\end{eqnarray}
We can now study the size distribution of short loops
at the high--energy regime, once equilibration is established.
We denote by $dn$ the number of loops with energies from $E$
to $E + dE$. Figure 2 shows
the energy distribution of short loops in the high--energy density
regime, as a function of space dimensionality. We have checked that
these distributions are independent of the particular value of $\rho$
($\rho >\rho_H$) for a given $D$.  We find that the size distribution of
short loops is well defined by a line
\begin{equation}
{dn\over dE}\ \ \sim \ \ E^{-(1+D/2)}\ , \label{distr}
\end{equation}
where the space dimensionality $D$ was taken equal to $3$, $4$, or $5$.
This is what was indeed expected from analytical string results.
The statistical errors indicate a slope equal to $-(1+D/2)\pm 0.2$.
Above the Hagedorn
energy density the system is characterized by a scale invariant
distribution of short loops and a number of long loops with a distribution
that is not scale invariant.

The evolution of a network of loops reveals the existence
of two relaxation times:  a relaxation time
$t_{rel}$ defined as the time by which the ratio $x_s$
equilibrates and a relaxation time $\tau_{rel}$ defined as
the time by which the slope in the size distribution of short
loops reached the value $-(1+D/2)$ and remained
constant thereafter. It is clear that while $t_{rel}$ has a
meaning for any value of string density, $\tau_{rel}$ is only
meaningful for string densities above the critical one $\rho_H$.
Both these  relaxation times are of the same order of magnitude,
which is at most about $100$ evolution timesteps.
The relaxation time seems to be independent of either the
string density $\rho $ or the space dimensionality $D$.

\section{Lifetime of Long Strings}

Our next goal is to study the lifetime of long loops
(which represent the winding modes) as a function of space
dimensionality and string energy density. As one expects, in the
high energy density regime, in other words above the Hagedorn
energy density, the long loops do remain indefinitely.
Around the critical energy density $\rho_H$, the lifetime
of long loops fluctuates a lot. The interesting regime is the
low energy density one, namely the case below the
Hagedorn energy density.  To address the issue of the lifetime
of long string loops in the low energy density regime,
as a function of space dimensionality, we basically pursue the
following approach:

In a $D$--dimensional toroidal box of size $L$, we create a
`loop gas' consisting of $s$--point loops
($s>2$), at a string energy density greater than the critical one,
but below the value of $\rho_H$ corresponding to $s=2$ (given
by Eq.~(\ref{critic}) for $D=3, 4, 5$).  [Note that, if for example
$D=3$, the critical string density goes roughly as $s^{-2}$.]
The energy of these loops is $E_s$.
We evolve the system until timestep $t_s$, under the condition that
the shortest loops which are allowed to be chopped off, must
have at least an energy $E_s$.
Since we are in the high--energy density regime, once the system reaches
equilibrium we expect the existence of
very long string modes, winding a number of times around the torus.
During the subsequent evolution timesteps, $t>t_s$,
we freeze out short loops up to an energy $E_f$, so that the rest
of the network has an energy density below the Hagedorn energy density
corresponding to $E_c=2$.
We then drop the energy cutoff to 2, which leads to an
increase of the critical string density $\rho_H$, reaching
the values given in Eq.~(\ref{critic}).
Now we are in the low--energy density regime and long string loops
would be expected to decay in order for equilibrium to be maintained.

In our simulations we create a network of twelve-- and two--point loops
at high enough energy densities, in boxes of size $L$ from 10 to 50 for
$D=3$ and $L$ from 10 to 20 for $D=4, 5$.  We evolve the loop string networks
until equilibrium is reached and long string states winding around the
toroidal box are formed, allowing loops of all sizes to reconnect, but
forbiding the creation of loops having less than 12 points.  We then
freeze small loops, up to usually six-- to eight--point loops, so that the
string density drops below the critical one for $E_c=2$.   We
subsequently set the cutoff $E_c$ to be indeed 2 and we see that while
for $D=3$ the long string loops decay after a few timesteps, for $D=4$ or
$D=5$, long loops representing the winding states  persist even after a
large number of evolution timesteps.  Figure 3 shows `snapshots' of the 3--$D$
string network (a) at $t=0$, when the system consists of twelve-- and
two--point loops; (b) at $t=t_s$, when the system consists of short loops
and long ones, winding around the box; and (c) at a relaxation time, showing
the decay of all winding string modes.

There are two other approaches which we followed in order to study the
lifetime of long string modes, and both of them support the previously stated
result.  In what follows, we will briefly describe those two methods.

1--Our initial configuration consists of a loop gas at a string
density $\rho $ above the critical value $\rho_H$, {\it i.e.,}
\begin{equation}
\rho \ \ =\ \ \alpha \ \rho_H \ \  (\alpha \ > \ 1)\ .
\end{equation}
We evolve this system and, once equilibrium is reached, the
density of long strings must be
\begin{equation}
\rho_l\ \ =\ \ (\alpha - 1)\ \rho_H \ .
\end{equation}
We are interested in the cases where $\rho_l$ is smaller than $\rho_H$.
(We have done various runs for $D=3$, $4$, or $5$ to verify the
above statement.) We can now disregard the short loops and consider the
network of only long loops. As a result of the imposed periodic boundary
conditions,  the total winding number is zero. We thus
evolve a very energetic string or
a network of a few very energetic strings in the low energy
density regime in a $D$--dimensional box, where $D=3$, $4$, or $5$.
The lifetime of winding modes is the inverse of the slope
of the $E_l$ versus time $t$ curve, where $E_l$ denotes
the number of lattice points (i.e., energy) occupied by long
loops. Our qualitative results indicate a strong dependence on
the space dimensionality $D$. While for $D=3$ the lifetime of
long loops is rather short (much shorter than the relaxation
time), as the space dimensionality increases ({\it i.e.,} $D > 3$),
the lifetime
of the very energetic loops increases dramatically;
winding modes remain `indefinitely'. At this point, one might
wonder whether the rather short lifetime of long loops, when the space
dimensionality $D=3$, is a result of the shape of those loops.
In other words, it might be that the very energetic loops survive
for a short time just because they are locally rather rough.
To check this possibility, we have studied the Hausdorff dimension
of loops. We estimate the Hausdorff dimension $\nu$ as follows:
we calculate the average distance, $d$, between two points on the
loop, as a function of the length, $l$, along the loop. The length was
found by connecting the points of the loop by straight lines, and the
averaging was done over a large number of loop segments having lengths
in specified narrow intervals (of unit width). The Hausdorff dimension
$\nu$ is defined as
\begin{equation}
\nu \ \ \equiv \ \ log (l)\ /\ log(d)\ .
\end{equation}
Our numerical simulations indicate that initially the
energetic strings are of
the form of random walks ($\nu = 2$) independently of space
dimensionality $D$. This is exactly what we were expecting
to find: since the very energetic strings were the outcome
of the evolution of a loop gas above the Hagedorn energy density,
their shape should indeed be approximately Brownian.
Now, as this low--energy density (short loops have been disregarded)
 network of winding modes evolves,
these long strings become almost straight (as it was indeed
expected).
We thus conclude that the disappearance
of winding modes in the case of $D = 3$ space dimensionality
is not the result of any particular local roughness
of the loops configurations.
The `weak' point of this approach is that the network of long string
states, which was created evolving the initial high--energy density
loop gas, consisted mainly of string loops winding only a few times
around the torus.

2--In a $D$--dimensional toroidal box of size $L$ we create a network of
long string states winding a large number of times around the torus.
We create this system in such a way that for each pair of opposite faces
 of the box, if a
string going from face $j$ towards face $j'$, cuts $n_j$--times this face
$j'$, then there must be another string state, having the opposite direction,
which cuts also $n_j$--times the face $j$. To this string configuration,
composed of a number of long string states winding many times around the
torus, we add zero--length links, so that we are always in the low--energy
density regime.  The evolution of such a string network shows that long
string states decay only if the space dimensionality of the torus is $D=3$.

The persistence of long loops at string energy
densities below the critical one, when the space dimensionality
is $D > 3$, supports the proposed cosmological scenario
by Brandenberger and Vafa \cite{br-v}, which attempts
to explain the space dimensionality and to avoid the initial
singularity by means of  string theory.
According to their scenario, the universe starts as a compact
space in a high number of space dimensions. Its expansion is
slowed down due to the existence of winding modes (their energy
grows linearly with $L$), which being
unable to find one another, fail to annihilate
and fall out of equilibrium. The universe will
contract until the momentum modes slow the contraction, as
the universe reached a size below the Planck scale (a consequence of the target
space duality). As a result the universe will undergo an oscillatory
period, avoiding both the initial singularity and the
divergence of the temperature at the Big Bang. At some point,
fluctuations might accidentally lead to a 3--dimensional expansion,
leading to the annihilation of winding modes and the irreversible
birth of the 3--dimensional decompactified universe \cite{br-v}. As those
authors suggest \cite{br-v}, in exactly $(3+1)$--spacetime, the 2--dimensional
world--sheets of the winding modes can intersect, unwinding
themselves. However, the fact that the maximum spacetime dimensionality
required in order to maintain thermal equilibrium is $4$, does not
explain why the space dimensionality is $3$ and not a lower one.
A suggestion to resolve this issue, which however
has not been worked out, was proposed in Refs.~\cite{br-v}, \cite{cat}.

\section{Conclusions}

In this paper we have studied via numerical experiments some classical
aspects of fundamental strings.  Our aim was basically to test a scenario of
string cosmology proposed by Brandenberger and Vafa \cite{br-v}, based on the
heterotic string theory in the space of a 9--dimensional toroidal universe of
 Planckian size and one time dimension.  The basic idea of the model is
that such a small universe underwent a long period of oscillations, until
finally three out of nine dimensions began to expand leading to our present
universe.  Performing a series of numerical simulations based on  a simple
and exact algorithm, we studied the thermal properties and the equilibration
process of a string network in a toroidal universe, and in particular its
dimension dependence.  Our results are in a full agreement with those found for
quantum strings.  At the low--energy density regime the string network is
dominated by the shortest allowd string loops, which represent the string
momentum modes.  As the energy density increases, a phase transition takes
place, characterized by the appearance of long string loops winding around
the torus; these represent the winding string modes.  The phase transition
is a generic property independent of the space dimensionality.  However the
space dimensionality plays a crucial role for the lifetime of the long string
loops.  As we found, only if the space dimensionality is equal to three, long
winding string modes decay in the low--energy density regime, while these modes
being unable to find one another, fail to annihilate and persist indefinitely
in a higher dimensional space.

\vspace{1cm}
{\bf Acknowledgement}
\vspace{0.5cm}\\
It is a pleasure to express my thanks to Cumrun Vafa, for raising my interest
in this problem, for a number of stimulating discussions which we had and for
his invitation to the Lyman Laboratory at Harvard University, where  part
of this work was done.  I would also like  to thank both Alex Vilenkin and
Robert Brandenberger for their suggestions, comments and helpful discussions.

\newpage
{\large Figure Captions}
\vspace{2cm}\\
{\bf Fig.~1a: }
The energy density of short loops, $\rho_s=\rho-\rho_l$, as
a function of the total string energy density $\rho$.  Strings are
moving in a 3--dimensional torus. The smallest
allowed loops consist of only two points.  There is
a sharp change of behaviour at $\rho_H\sim 0.170$.
\vspace{1cm}\\
{\bf Fig.~1b: }
The energy density of short loops, $\rho_s=\rho-\rho_l$, as
a function of the total string energy density $\rho$.  Strings are
moving in a 4--dimensional torus. The smallest
allowed loops consist of only two points.  There is
a sharp change of behaviour at $\rho_H\sim 0.062$.
\vspace{1cm}\\
{\bf Fig.~1c: }
The energy density of short loops, $\rho_s=\rho-\rho_l$, as
a function of the total  string energy density $\rho$.  Strings are
moving in a 5--dimensional torus. The smallest
allowed loops consist of only two points.  There is
a sharp change of behaviour at $\rho_H\sim 0.031$.
\vspace{1cm}\\
{\bf Fig.~2: }
Energy distributions of short loops in the high--energy
density regime, for strings moving in a $D$--dimensional torus with
$D=3, 4, 5$, as we go from right to left respectively.
These distributions do not depend on the particular value of $\rho$
($\rho > \rho_H$) and are well fitted by $dn/dE\ \sim \ E^{-(1+D/2)}$.
\vspace{1cm}\\
{\bf Fig.~3a.1: }
Snapshot of the initial state of a string gas (a) in
a 3--dimensional torus.  It consists of 12-- and 2--point loops
in a box of $L=30$, where 5400 points are occupied by strings.
\vspace{1cm}\\
{\bf Fig.~3a.2: }
Snapshot of the string system (a) at $t=60$. Only the
 2 long strings having energies equal to 417, 1113 respectively, are shown.
Until this timestep the smallest allowed energy cutoff is $E_c=12$.
\vspace{1cm}\\
{\bf Fig.~3a.3: }
Snapshot of the string system (a) at the relaxation time
$t=109$.  There are no long string modes and they do not appear later on.
The cutoff was dropped to $E_c=2$, while $E_f=6$.
\vspace{1cm}\\
{\bf Fig.~3b.1:}
Snapshot of the initial state of a string gas (b) in
a 3--dimensional torus.  It consists of 12-- and 2--point loops
in a box of $L=20$, where 2200 points are occupied by strings.
\vspace{1cm}\\
{\bf Fig.~3b.2: }
Snapshot of the string system (b) at $t=100$. Only the
 4 long strings having energies equal to 168, 238, 255, 471
respectively, are shown.  Until this timestep the smallest allowed
energy cutoff is $E_c=12$.
\vspace{1cm}\\
{\bf Fig.~3b.3: }
Snapshot of the string system (b) at the relaxation time
$t=153$.  There are no long string modes and they do not appear later on.
The cutoff was dropped to $E_c=2$, while $E_f=8$.


\begin{thebibliography}{99}
\bibitem{n} Y. Nambu, in proceedings of Int. Conf. on {\em Symmetries and
Quark Models}, (Wayne State University) (1969); Lectures at the Copenhagen
Summer Symposium (1970).
\bibitem{g} T. Goto, {\em Prog. Theor. Phys.} {\bf 46}, 1560 (1971).
\bibitem{shell} P. Shellard, {\em Nucl. Phys.} B{\bf 283}, 624 (1987).
\bibitem{alex}  A.G. Smith and A. Vilenkin, {\em Phys. Rev.}
D{\bf 36}, 990 (1987).
\bibitem{br-v} R. Brandenberger and C. Vafa, {\em Nucl. Phys.}
B{\bf 316}, 391 (1989).
\bibitem{dual} K. Kikkawa and M. Yamasaki, {\em Phys. Lett.}
B{\bf 149}, 357 (1984); V. Nair, A. Shapere, A. Strominger and
F. Wilczek, {\em Nucl. Phys.} B{\bf 287}, 402 (1987);
P. Ginsparg and C. Vafa, {\em Nucl. Phys.}
B{\bf 289} 414 (1987); B. Sathiapalan, {\em Phys. Rev. Lett.} {\bf 58},
1597 (1987); A.A. Tseytlin, {\em Mod. Phys. Lett.} A{\bf 6}, 1721 (1991).
\bibitem{ts-v} A.A Tseytlin and C. Vafa, {\em Nucl. Phys.} B{\bf 372},
443 (1992).\bibitem{bw} M.J. Bowick and L.C.R. Wijewardhana, Phys. Rev.
Lett. {\bf 54}, 2485 (1985).
\bibitem{nt} D. Mitchell and N. Turok, {\em Phys. Rev. Lett.} {\bf 58},
1577 (1987); {\em Nucl. Phys.}  B{\bf 294}, 1138 (1987).
\bibitem{jain} N. Deo, S. Jain and C. Tan, {\em Phys. Rev.} D{\bf 40},
2626 (1989); {\em Phys. Lett.} B{\bf 220}, 125 (1989).
\bibitem{old} P. Goddard, J. Goldstone, C. Rebbi and C.B.
Thorn, {\em Nucl. Phys.} B{\bf 56}, 109 (1973).
\bibitem{hag} R. Hagedorn, {\em Nuovo Cimento Suppl.} {\bf 3}, 147 (1965).
\bibitem{carl} R.D. Carlitz, {\em Phys. Rev.} D{\bf 5}, 3231 (1972).
\bibitem{fr} S. Frautschi, {\em Phys. Rev.} D{\bf 3}, 2821 (1971).
\bibitem{mairi1} M. Sakellariadou and A. Vilenkin, {\em Phys. Rev.}
D{\bf 37}, 885 (1988).
\bibitem{vavi} T. Vachaspati and A. Vilenkin, {\em Phys. Rev.} D{\bf 30},
2036 (1984).
\bibitem{cat} H. C\^ateau and K. Sumiyoshi, Phys. Rev. D {\bf 46},
2366 (1992).
\end{thebibliography}
\end{document}